\documentclass[a4paper,11pt]{article}

\usepackage{jheppub} 
\usepackage[T1]{fontenc}
\usepackage[utf8]{inputenc}
\usepackage[english]{babel}
\usepackage{amsmath,bbm}
\usepackage{amsfonts}
\usepackage{amssymb}
\usepackage{graphicx}
\usepackage[normalem]{ulem}
\usepackage{tabularx}
\usepackage{hyperref}
\hypersetup{unicode,psdextra}
\usepackage{scalerel}
\usepackage{mathabx} 
\usepackage{cancel}
\usepackage{ytableau}

\newcommand*{\eweakgroup}{\mbox{$\mathit{SU}(2)_L \times U(1)_Y$}}

\newcommand*{\matheweakgroup}{\mathit{SU}(2)_L \times U(1)_Y}
\newcommand*{\mathemgroup}{U(1)_{em}}
\newcommand*{\sutwo}{$\mathit{SU}(2)_L$}

\newcommand*{\unitmatrix}{\mathbbm{1}}
\newcommand*{\twomat}[1]{\underline{#1}}             
\newcommand*{\tvec}[1]{\boldsymbol{#1}}              

\newcommand*{\trans}{\mathrm{T}}                     

\DeclareMathOperator{\diag}{diag}		
\DeclareMathOperator{\tr}{Tr}		
\DeclareMathOperator{\rank}{rank}		

\title{Electroweak gauge invariant Higgs multiplets}

\author{M.~Maniatis}

\affiliation{Centro de Ciencias Exactas,  Universidad del B\'io-B\'io,  Casilla  447,  Chill\'{a}n, Chile}

\emailAdd{maniatis8@gmail.com}
 
 \abstract{We investigate the potential of general Higgs multiplets. 
We establish the domain of general Higgs multiplets within the context of the adjoint representation.
The construction of the most general potential for arbitrary multiplets is achieved using the irreducible, totally symmetric representation of tensor products of doublets. Furthermore, we explore symmetry transformations of the Higgs multiplets, with particular emphasis on CP symmetries.
 }
\begin{document} 

\maketitle
\flushbottom
\section{Introduction}
T.D. Lee originally proposed extending the Higgs sector to two Higgs-boson doublets to provide a new source of CP violation~\cite{Lee:1973iz}. This motivation is still today valid. For a review of the two-Higgs doublet model~(2HDM) see for instance~\cite{Branco:2011iw}. The 2HDM has been studied with respect to stability, electroweak symmetry breaking, CP symmetries, as well as general symmetries; see for instance~\cite{Gunion:1989we,Nishi:2006tg,Maniatis:2007vn,Degee:2009vp,Ferreira:2010yh,Arhrib:2010ju,Aoki:2011wd,Barroso:2013awa,Ginzburg:2016tlh,Basler:2016obg,Grzadkowski:2016szj,Bento:2020jei,Kanemura:2022cth,Heinemeyer:2024hxa}. 

Most extensions of Higgs multiplets in the literature focus on singlet and doublet copies.
In supersymmetric models, for instance, additional Higgs-boson singlets and doublets appear in a rather natural way.
As an example let us mention the next-to minimal supersymmetric model - see the reviews~\cite{Maniatis:2009re, Ellwanger:2009dp} - which besides two Higgs-boson doublets contains one complex singlet. 

What about Higgs-boson triplets, quadruplets, and so on? With respect to general multiplets there is a severe restriction arising from the parameter defined as
\begin{equation} \label{rhoex}
\rho = \frac{m_W^2}{\cos^2 (\theta_W) m_Z^2}\;.
\end{equation}
Here the masses of the electroweak gauge bosons are denoted by $m_W$ and $m_Z$ and the electroweak mixing angle by~$\theta_W$. This parameter is measured to be close to one~\cite{ParticleDataGroup:2022pth}.
As usual, from the kinetic terms of the Higgs multiplets, written gauge covariantly, we compute the masses of the electroweak gauge bosons. From the neutral components of the Higgs multiplets we get contributions to the masses of the electroweak gauge bosons. Numbering the Higgs multiplets with $i$, denoting their weak isospin by~$T(i)$, its $3$ component by $T_3(i)$, and the vacuum-expectation value of its neutral component by~$v_i$, the $\rho$ parameter at tree level is given by~\cite{ParticleDataGroup:2022pth},
\begin{equation} \label{rhoth}
\rho =
\frac{\sum_i \left( T(i) (T(i)+1) - T_3^2(i) \right) |v_i|^2 }
{2 \sum_i T_3^2(i)  |v_i|^2}\;.
\end{equation}
From this computation of the $\rho$ parameter follows that 
for Higgs-boson singlet models with $T=T_3=0$ and Higgs-boson doublet models with $T=1/2$ and $T_3=\pm1/2$ we find 
$\rho=1$, consistent with experimental observation. One special case is the Standard Model which, since it employs one Higgs-boson doublet, predicts also $\rho=1$ at tree level.
 
An example of a higher multiplet resulting in $\rho=1$, as follows from~\eqref{rhoth}, are Higgs septuplets corresponding to $T=3$ and $T_3= \pm 2$. Another possibility to comply with this constraint is to add multiplets with vanishing vacuum-expectation values of their neutral component. Also it is possible to have a potential with several multiplets and with the vacuum-expectation values arranged by an appropriate potential to eventually give $\rho =1$.
Examples are the Georgi-Machacek models~\cite{Georgi:1985nv, Chanowitz:1985ug} with one Higgs-boson doublet besides Higgs-boson triplets. The potential is arranged in these type of models such that the vacuum-expectation values satisfy at tree level the condition~$\rho=1$. Models with additional multiplets with small vacuum-expectation values are considered for instance in~\cite{Giarnetti_2024}, giving raise to neutrino masses.
In~\cite{Jurciukonis:2024cdy} a model of a doublet with an additional multiplet with weak isospin up to $T=7/2$ is considered, which does not develop a vacuum-expectation value. 
 Let us also mention the recent works~\cite{Kannike:2023bfh, Jurciukonis:2024bzx}, where a Higgs-boson quadruplet is combined with a doublet.
Eventually, even with~\eqref{rhoth} in mind, we see that it is at least in principle possible to consider models with more general Higgs-boson multiplets. In this sense, we here would like to study Higgs potentials of arbitrary Higgs multiplets. 

Stability and electroweak symmetry breaking of Higgs-multiplet models have been studied in~\cite{Abud:1981tf} and \cite{Kim:1981xu}; for the case of an arbitrary number of doublets in~\cite{Maniatis:2015gma}. Couplings of Higgs multiplets to the electroweak gauge bosons have been
investigated in~\cite{Bernreuther:1998rx}. 
 
Let us briefly return to the Standard Model, where a single Higgs-boson doublet field~$\varphi^{(2)}(x)$ belongs to 
$\mathbbm{C}^2$. This field transforms as a doublet under electroweak \sutwo{} transformations as indicated by the superscript. It is straightforward to introduce a {\em bilinear} field,
\begin{equation}
K^{(2)}(x) = {\varphi^{(2)}}^\dagger(x) \varphi^{(2)}(x)
\end{equation}
which is bilinear in the doublet fields, real, and it is a singlet under electroweak gauge transformations. Bilinear fields have been studied in detail for the case of an arbitrary number of doublets~\cite{Maniatis:2015gma}. Here we will generalize the bilinear approach to arbitrary multiplets. 
Specifically, the bilinear representation enables us to determine the domain of these multiplets.
This domain can be split into distinct parts with respect to  the behavior of the multiplets with respect to electroweak symmetry breaking, that is, unbroken, fully, and partially broken electroweak symmetry $\matheweakgroup \to \mathemgroup$.

Having found the domain of the multiplets with respect to electroweak-symmetry breaking, we proceed and study the corresponding Higgs potential. 
In the potential we shall consider multiplets in the adjoint, that is, bilinear representation. 
This has many advantages, for instance with respect to the study of symmetries of the potential. For instance, for the case of two Higgs-boson doublets, it has been shown that CP transformations are given by reflections in terms of bilinears~\cite{Nishi:2006tg,Maniatis:2007vn}. Let us mention in this context that for the case of the 2HDM, a description free of gauge redundancies has been given for the complete model~\cite{Sartore:2022sxh}, including the Yukawa and gauge couplings. 

However, considering potentials of different multiplets it turns out to be possible to form gauge-invariant terms which are not bilinear in all the multiplets. One example is a gauge-invariant tensor product of a pair of Higgs-boson doublets with one Higgs-boson triplet. Motivated by a recent study of Higgs-boson quadruplets and doublets~\cite{Kannike:2023bfh}, 
we show how these gauge-invariant terms can be formed systematically. The essential idea is to express the multiplets in terms of irreducible symmetric doublet representations.

Finally, we use the description of Higgs-boson multiplets in both the adjoint and symmetric representations to examine the general symmetries of the potential.
As an example we study standard CP symmetries for the case of general Higgs multiplets.

\section{Higgs multiplets}

We are considering Higgs-boson multiplets with respect to electroweak gauge symmetry. In analogy with ordinary spin regarding the group \sutwo, the Higgs multiplets have to have isospin $T=0, 1/2, 1, 3/2, 2, \ldots$ with the 3 component running from $T_3 = -T, -T+1, \ldots, +T$. Considering the group $U(1)_Y$,  each Higgs multiplet is assigned a hypercharge~$Y$. 
\begin{table}[h!]
\begin{center}
\begin{tabular}{lccccc}
\hline
isospin $T$ & $0$ & $1/2$ & $1$& $3/2$ & $\frac{m-1}{2}$\\
$T_3$ & $0$ & $-1/2, +1/2$ & $-1, 0, +1$ & $-3/2, \ldots, 3/2$ & ${\scriptstyle -\frac{m-1}{2}}, \ldots, {\scriptstyle +\frac{m-1}{2}}$\\
multiplicity & $1$ & $2$ & $3$ & $4$ & $m$\\
fund. repr. $\varphi^{(m)}$ & 
$\phi^1$ & 
$\begin{pmatrix} \phi^1\\ \phi^2\end{pmatrix}$ & 
$\begin{pmatrix} \phi^1\\ \phi^2\\ \phi^3\end{pmatrix}$  &
$\begin{pmatrix} \phi^1\\ \vdots\\ \phi^4\end{pmatrix}$  &
$\begin{pmatrix} \phi^1\\ \vdots\\ \phi^m \end{pmatrix}$\\
\sutwo{} tensor product &
$\Delta$ & 
$\Delta_{i}$ &
$\Delta_{(ij)}$ &
$\Delta_{(ijk)}$ &
$\Delta_{(i_1,\ldots, i_{m-1})}$\\
\hline
\end{tabular}
\end{center}
\caption{\label{tabmul}Higgs multiplets and their properties. For different multiplets with multiplicity $m$ the isospin~$T$, its $3$ components $T_3$, the multiplets in the fundamental representation and the irreducible \sutwo{}
representation which is totally symmetric in the corresponding indices $i,j,k, i_1, \ldots, i_{m-1} \in \{1,2\}$ are shown.
}
\end{table}
The Higgs multiplets corresponding to isospin~$T$ have multiplicity $m= 2T+1$. The charges of the components of the Higgs fields in the fundamental representation follow from the Gell-Mann-Nishijima formula $Q = Y + T_3$~\cite{ParticleDataGroup:2022pth}. If we choose, for instance, a hypercharge $Y= -T_3$ we find the lowest component of the Higgs-multiplet in the fundamental representation, corresponding to $-T_3$, to be electrically neutral. This of course is only one example and any hypercharge may be assigned to the multiplets. Independent of the assignment of the hypercharge the electric charge decreases from top to bottom in each component by one elementary charge unit.

In general, the tensor product of $m-1$ copies of doublets of \sutwo, that is, $2 \otimes 2 \otimes \ldots \otimes 2$ has one totally symmetric irreducible representation of dimension~$m$.  It is very convenient to represent all multiplets in the symmetric representation in order to form gauge-invariant tensor products among them. 

Let us examine the totally symmetric irreducible representations
 $\Delta_{(a_1, \ldots, a_{m-1})}$ in detail. As usual, the round brackets indicate that the tensor is totally symmetric under any permutation of the indices. 
For instance, the \sutwo{} Higgs triplet corresponding to multiplicity $m=3$ can be represented as,
\begin{equation}
\Delta_{11} = \phi^1, \qquad
\Delta_{12} = \Delta_{21} = \frac{1}{\sqrt{2}} \phi^2, \qquad
\Delta_{22} = \phi^3\;.
\end{equation}
If we assign to the triplet the hypercharge~$Y=1$
we get with the help of the Gell-Mann-Nishijima formula the expressions
\begin{equation}
\Delta_{11} = \phi^{++}, \qquad
\Delta_{12} = \Delta_{21} = \frac{1}{\sqrt{2}} \phi^{+}, \qquad
\Delta_{22} = \phi^{(0)}\;,
\end{equation}
that is,
\begin{equation}
\left(\Delta_{ij}\right) = 
\begin{pmatrix}
\phi^{++} & \frac{1}{\sqrt{2}} \phi^{+}\\
\frac{1}{\sqrt{2}} \phi^{+} & \phi^{(0)}
\end{pmatrix}.
\end{equation}
Similar, we find for a quadruplet in terms of the \sutwo{}
symmetric representation the explicit expressions,
\begin{multline}
\Delta_{111} = \phi^{1}, \qquad
\Delta_{112} = \Delta_{121} = \Delta_{211}  = 
\frac{1}{\sqrt{3}} \phi^{2}, \\
\Delta_{122} = \Delta_{212} = \Delta_{221}  = 
\frac{1}{\sqrt{3}} \phi^{3}, \qquad
\Delta_{222} = \phi^{4}\;.
\end{multline}
In general, the totally symmetric doublet product representation for a multiplet of multiplicity~$m$ in terms of the fundamental components reads
\begin{equation}
\Delta_{(\underbrace{1, \ldots, 1}_{n_1}, 
\underbrace{2, \ldots, 2}_{m-n_1-1})}
=
\frac{1}{\sqrt{\binom{m-1}{n_1}}} \phi^{m-n_1}.
\end{equation}
Since the doublet representation is totally symmetric as indicated by the round brackets enclosing the indices, this expression holds for all permutations of the indices 1 and 2. 
In Tab.~\ref{tabmul} we summarize the properties of the Higgs multiplets. 
\\

Now we seek to determine the domain of the Higgs multiplets - avoiding systematically gauge redundancies. 
To achieve this we construct the Higgs multiplets in the adjoint representation in an analogous way as outlined in~\cite{Maniatis:2006fs} for the case of Higgs-boson doublets. The case of three Higgs-boson doublets, the 3-Higgs doublet model (3HDM), has been studied in detail in~\cite{Maniatis:2014oza} and in~\cite{Maniatis:2015gma} the case of an arbitrary number of doublets, the $n$-Higgs doublet model (nHDM), is discussed. 
Let us denote by $n_m$ the number of identical copies of Higgs-boson multiplets with multiplicity~$m$. In particular $n_1$  denotes the number of singlets, which carry electroweak isospin $T=0$, that is, $T=0=T_3=0$. Of course, singlets are by definition gauge invariant. 
More generally, for the case of $n_m$ copies of Higgs bosons of multiplicity $m$ we build the matrix 
\begin{equation} \label{psi2}
\psi^{(m)} = \begin{pmatrix} {\varphi_1^{(m)}}^\trans\\ \vdots\\ {\varphi_{n_m}^{(m)}}^\trans \end{pmatrix}
=
\begin{pmatrix} 
\phi^1_1& \cdots  & \phi^m_1\\
\vdots & \vdots\\
\phi^{1}_{n_m}& \cdots & \phi^m_{n_m}
\end{pmatrix}
 \in (n_m \times m)\;.
\end{equation}
Note that for the entries of this matrix, for instance~$\phi^2_3$, the subscript labels the number of the copy, whereas the superscript gives the component of this copy of the multiplet.

For arbitrary multiplets the next step is to form from $\psi^{(m)}$ the gauge-invariant matrix: 
\begin{equation} \label{Kmatdef}
\twomat{K}^{(m)} = \psi^{(m)} {\psi^{(m)}}^\dagger =
\begin{pmatrix} 
{\varphi_1^{(m)}}^\dagger \varphi_1^{(m)} & \cdots & {\varphi^{(m)}_{n_m}}^\dagger \varphi^{(m)}_1 \\
\hdots & \ddots & \hdots\\
{\varphi_1^{(m)}}^\dagger \varphi^{(m)}_{n_m} & \cdots & {\varphi^{(m)}_{n_m}}^\dagger \varphi^{(m)}_{n_m}
\end{pmatrix} \in (n_m \times n_m)\;.
\end{equation}
By construction, the quadratic, gauge-invariant matrix $\twomat{K}^{(m)}$ is hermitian, has rank smaller than or equal to~$m$ and is positive semidefinite. The rank condition follows from the fact that the $(n_m \times m)$ matrix $\psi^{(m)}$ has trivially a rank smaller than or equal to~$m$, whereas hermiticity and semi definiteness follow directly from the definition of the matrix~$\twomat{K}^{(m)}$.

As an example let us consider the two-Higgs triplet model. The two triplets correspond to $T=1$ with multiplicity $m=3$. We have in this case
\begin{equation} \label{psi3ex}
\psi^{(3)} = \begin{pmatrix} {\varphi^{(3)}_1}^\trans\\ {\varphi^{(3)}_{2}}^\trans \end{pmatrix} =
\begin{pmatrix}
\phi^{1}_1  & \phi^2_1 & \phi^3_1\\
\phi^{1}_2  & \phi^2_2 & \phi^3_2
\end{pmatrix}.
\end{equation}
If we assign the hypercharge $Y=1$ to these two triplets we will get the charges of the components,
\begin{equation}
\psi^{(3)} = \begin{pmatrix} {\varphi^{(3)}_1}^\trans\\ {\varphi^{(3)}_{2}}^\trans \end{pmatrix} =
\begin{pmatrix}
\phi^{++}_1  & \phi^+_1 & \phi^0_1\\
\phi^{++}_2  & \phi^+_2 & \phi^0_2
\end{pmatrix}.
\end{equation}
The matrix $\twomat{K}^{(3)}$ defined in~\eqref{Kmatdef} for the case~$\psi^{(3)}$ in~\eqref{psi3ex} is of dimension $2 \times 2$ and reads
\begin{equation}
\twomat{K}^{(3)} = \psi^{(3)} {\psi^{(3)}}^\dagger =
\begin{pmatrix} 
{\varphi^{(3)}_1}^\dagger \varphi^{(3)}_1\;\;\; & {\varphi^{(3)}_2}^\dagger \varphi^{(3)}_1 \\
{\varphi^{(3)}_1}^\dagger \varphi^{(3)}_2\;\;\; & {\varphi^{(3)}_2}^\dagger \varphi^{(3)}_2
\end{pmatrix}.
\end{equation}

The next step is to write the gauge-invariant bilinear matrix $\twomat{K}^{(m)}$ for each set of the $n_m$ multiplets in a basis of the unit matrix and the (generalized) Pauli matrices,
\begin{equation} \label{twomatKbasis}
\twomat{K}^{(m)} = \frac{1}{2} K^{(m)}_\alpha \lambda_\alpha, \qquad \alpha \in \{0, \ldots, n_m^2-1\}\;.
\end{equation}
Here we have added to the generalized Pauli matrices the conveniently scaled identity matrix written with an index zero,
 \begin{equation} \label{idscaled}
 \lambda_\alpha \text{ with }  \alpha \in \{0, \ldots, n_m^2-1\}, \qquad
 \lambda_0 = \sqrt{\frac{2}{n_m}} \unitmatrix_{n_m}\;.
 \end{equation}
 An explicit scheme to construct the generalized Pauli matrices can for instance be found in~\cite{Maniatis:2015gma}. 
 The matrices $\lambda_\alpha$ satisfy the equations,
 \begin{equation} \label{lambdaeq}
 \tr(\lambda_\alpha \lambda_\beta) = 2 \delta_{\alpha \beta}, \qquad
 \tr(\lambda_\alpha) = \sqrt{2 n_m} \delta_{\alpha 0 }\;.
 \end{equation}
 Multiplying~\eqref{twomatKbasis} on both sides with (generalized) Pauli matrices and taking traces we can express the bilinears in terms of the Higgs multiplets,
 \begin{equation} \label{bilinearsexp}
 K^{(m)}_\alpha = \tr ( \twomat{K}^{(m)} \lambda_\alpha)\;.
 \end{equation}
 These bilinears are real, gauge invariant expressions. 

 In our example of two Higgs triplets we find for the real gauge-invariant bilinears explicitly
 \begin{equation} \label{Kexample}
 \begin{split}
 &K^{(3)}_0 = {\varphi^{(3)}_1}^\dagger \varphi^{(3)}_1 + {\varphi^{(3)}_2}^\dagger \varphi^{(3)}_2, \\
 &K^{(3)}_1 = {\varphi^{(3)}_1}^\dagger \varphi^{(3)}_2 + {\varphi^{(3)}_2}^\dagger \varphi^{(3)}_1, \\
 &K^{(3)}_2 = i {\varphi^{(3)}_2}^\dagger \varphi^{(3)}_1 - i {\varphi^{(3)}_1}^\dagger \varphi^{(3)}_2, \\
 &K^{(3)}_3 = {\varphi^{(3)}_1}^\dagger \varphi^{(3)}_1 - {\varphi^{(3)}_2}^\dagger \varphi^{(3)}_2\;.
 \end{split}
 \end{equation}

Let us determine the domain of the real bilinears $K^{(m)}_\alpha$, $\alpha \in \{0, \ldots, n_m^2-1\}$. 
From the explicit construction we get for the zero component
\begin{equation} \label{Km0}
K^{(m)}_0 =  \tr ( \twomat{K}^{(m)} \lambda_0) 
   = \sqrt{\frac{2}{n_m}} 
	\left(
 {\varphi_1^{(m)}}^\dagger \varphi^{(m)}_1 + \ldots +  {\varphi^{(m)}_{n_m}}^\dagger \varphi^{(m)}_{n_m}
 	\right)
 \end{equation} 
  and we find immediately 
 \begin{equation} \label{Km0ge0}
 K^{(m)}_0 \ge 0\;.
 \end{equation}
 The rank condition $\rank(\twomat{K}^{(m)}) \le m$ translates to the bilinears as follows (here we follow closely~\cite{Maniatis:2015gma}):
 Since $\psi^{(m)}$ has trivially a rank smaller than or equal to $m$, this holds also for the bilinear matrix
 $\twomat{K}^{(m)}=\psi^{(m)} {\psi^{(m)}}^\dagger $. We can diagonalize $\twomat{K}^{(m)}$ by a unitary similarity transformation,
 \begin{equation} \label{diagKm}
 U \twomat{K}^{(m)} U^\dagger = \diag(\kappa_1, \ldots, \kappa_m, 0, \ldots, 0)\;.
 \end{equation}
 Positive semidefiniteness translates for the eigenvalues to
 \begin{equation}
 \kappa_i \ge 0, \qquad i \in \{1, \ldots, m\}\;.
 \end{equation}
 For any hermitean matrix $\twomat{K}^{(m)} \in (n_m \times n_m)$ with eigenvalues $\kappa_i$ we can introduce the symmetric sums,
 \begin{equation} \label{defsymmsum}
 s_0 = 1, \qquad s_k = \sum\limits_{1 \le i_1 < i_2 < \ldots < i_k \le n_m}
 \kappa_{i_1} \kappa_{i_2} \cdots \kappa_{i_k}, \qquad k \in \{1, \ldots, n_m\}\;.
 \end{equation}
 In particular we have $s_0 = 1$, $s_1 = \kappa_1 + \cdots + \kappa_{n_m}$ and
 $s_{n_m} = \kappa_1 \cdot \ldots \cdot \kappa_{n_m} = \det(\twomat{K}^{(m)})$. 
 The symmetric sums can be expressed recursively in terms of basis-independent traces of powers of the bilinear matrix~\cite{Maniatis:2015gma},
 \begin{equation} \label{ssumK}
 s_0 = 1, \qquad
 s_k = \frac{1}{k} \sum\limits_{i=1}^k (-1)^{i-1} s_{k-i} \tr( (\twomat{K}^{(m)})^i), \quad k \in \{1, \ldots, n_m\}\;.
 \end{equation}
 Explicitly, the first symmetric sums read
 \begin{equation} \label{ssymsexp}
 \begin{split}
 &s_0 = 1,
 \\
 &s_1 = \tr(\twomat{K}^{(m)}) = \sqrt{\frac{n_m}{2}} K^{(m)}_0, 
 \\
 &s_2 = \frac{1}{2} \left( \tr^2(\twomat{K}^{(m)}) - \tr((\twomat{K}^{(m)})^2) \right),
 \\
 &s_3 = \frac{1}{6} \left( \tr^3(\twomat{K}^{(m)}) - 3 \tr((\twomat{K}^{(m)})^2)\tr(\twomat{K}^{(m)}) + 2 \tr((\twomat{K}^{(m)})^3) \right)\;.
 \end{split}
 \end{equation}
As stated before, the matrix $\twomat{K}^{(m)}$ has rank smaller than or equal to $m$.  Having $\twomat{K}^{(m)}$ of rank
 $r$, with $0 \le r \le m$, this is equivalent to the first $r$ symmetric sums positive and the remaining ones vanishing. This can be shown as follows:
Suppose we have
a matrix $\twomat{K}^{(m)}$ of rank $r$, $0 \le r \le m$. The number of positive, non-vanishing eigenvalues is therefore $r$ and without loss of generality we set
\begin{equation} \label{kappacond}
\kappa_i > 0 \quad \text{for } i \in \{1, \ldots, r\}, \qquad
\kappa_{r+1} = \kappa_{r+2}= \ldots = \kappa_{n_m} = 0\;.
\end{equation}
From the definition of the symmetric sums in~\eqref{defsymmsum} we see that $s_k = 0$ for $k>r$ since each summand in the sum is formed from a product of $k$ different eigenvalues, where at least one is vanishing. 
Therefore, for a matrix $\twomat{K}^{(m)}$ with rank $0 \le r \le m$ we have
\begin{equation} \label{ssumscond}
s_i  >  0 \quad \text{for } i \in \{1, \ldots, r\}, \qquad
s_{r+1} = s_{r+2}= \ldots = s_{n_m} = 0\;.
 \end{equation}
 
Vice versa, suppose we have the conditions for the symmetric sums as given in~\eqref{ssumscond}. 
From the definition of the sums we see that starting with the last symmetric sum condition, $s_{n_m}=0$, that at least one eigenvalue, say $\kappa_{n_m}$, without loss of generality, has to vanish. Recursively from the preceding condition we find that successively one more eigenvalue has to vanish. Eventually we find all the equality conditions in~\eqref{kappacond} to be satisfied.
It remains to be shown that for $s_i > 0$ with $i \in \{1, \ldots, r\}$ follows $\kappa_i >0$. This can be proven by induction.
The induction base follows from considering $r=1$: For $s_1>0$ we have $\kappa_1 = s_1>0$ and this forms the base of induction. Now we assume that for $s_i >0$, $i \in \{1, \ldots, r\}$ it follows that $\kappa_i >0$. Under this assumption we have to show that this holds for $r+1$. We can see this by considering the last condition, that is, $s_{r+1} = \kappa_1 \cdot \ldots \cdot \kappa_r \cdot \kappa_{r+1} > 0$. Under the assumption $\kappa_1 \cdot \ldots \cdot \kappa_r >0$ follows 
$\kappa_{r+1} > 0$, which completes the proof.

 With these preparations we can now find the domain of the bilinears. Each Higgs multiplet of multiplicity $m$ consists of $m$ complex components, therefore, the domain is $\varphi^{(m)} \in \mathbbm{C}^m$. 
 Of course, the domain is not unique with respect to gauge symmetries.
 For the general case, where we have $n_m$ Higgs multiplets of multiplicity $m$, we have the domain $\psi^{(m)} \in \mathbbm{C}^{n_m \times m}$. 
 The bilinears on the other hand are constructed from the bilinear matrix, that is, $\twomat{K}^{(m)} = \psi^{(m)} {\psi^{(m)}}^\dagger$ of dimension $n_m \times n_m$ with rank smaller than or equal to $m$. Therefore, the domain of the bilinears with respect to $n_m$ Higgs multiplets of multiplicity $m$ is given in terms of the symmetric sums $s_i$ from~\eqref{ssumK},
 \begin{equation} \label{domK}
 s_i \ge 0 \quad \text{for } i \in \{1, \ldots, m\}\;, 
 \qquad
 s_i = 0 \quad \text{for } i \in \{m+1, \ldots, n_m\}\;.
 \end{equation}
 The expressions for the symmetric sums in terms of bilinears are given in~\eqref{ssumK}. In this way we have found the domain of the multiplets in a gauge-invariant way.
 Let us see this in some examples: For $n_1$ Higgs complex singlets corresponding to multiplicity $m=1$, that is,
 $\varphi^{(1)}_1, \ldots,  \varphi^{(1)}_{n_1}$ we have $\psi^{(1)}$ of rank smaller than or equal to one. Therefore, in terms of bilinears, we have $s_1 \ge 0$ that is $K^{(1)}_0 \ge 0$ as well as $s_2=0, \ldots, s_{n_1}=0$. 
 The case of doublets has been discussed in terms of bilinears in~\cite{Maniatis:2006fs, Maniatis:2014oza, Maniatis:2015gma}. Let us recall the results. In case of $n_2$ Higgs-boson doublets, we have to have $\twomat{K}^{(2)}$ of rank smaller than or equal to two. This gives $s_1= K_0 \ge 0$ as well as $s_2 \ge 0$ and $s_3= \dots = s_{n_2} =0$. For the special case $n_2=2$, that is, the 2HDM we have, for instance, the domain of the bilinears $K^{(2)}_0 \ge 0$, 
 $s_2= \frac{1}{2} (\tr^2(\twomat{K}^{(2)}) - \tr((\twomat{K}^{(2)})^2)) = 1/4\big[
 (K^{(2)}_0)^2 - (K^{(2)}_1)^2 -(K^{(2)}_2)^2-(K_3^{(2)})^2 \big] \ge 0$. Note that we can always express the entries of the bilinear matrix $\twomat{K}^{(m)}$ in terms of bilinear coefficients $K^{(m)}_\alpha$. 

Having established the rank conditions for the bilinears it eventually remains to be shown that there is a one-to-one map of the Higgs multiplets in the fundamental representation onto the multiplets in the adjoint, that is, bilinear, representation - except for gauge redundancies. In Appendix~\ref{onetoone} this is shown analogously to the case of Higgs-boson doublets~\cite{Maniatis:2006fs}.

Let us recall that the Higgs-boson multiplets play a special role in particle physics since they provide in particular the masses to the fermions. In the Standard Model, for instance, one Higgs-boson doublet provides the masses to all fermions. In the two-Higgs doublet model, 2HDM, where the matrix $\twomat{K}^{(2)}$ is of rank smaller than or equal to two, the observed electroweak symmetry breaking with a neutral vacuum of the doublets singles out the rank one case. The rank two case correspond to a completely broken electroweak symmetry and the rank zero case corresponds to an unbroken electroweak symmetry. With view at the Gell-Mann-Nishijima formula, we see that 
 for any assignment of hypercharge to a multiplet in the fundamental representation the electric charge drops by one unit from top to down of the components. This means that we can have at most one neutral component of the Higgs-boson multiplets. In case only the neutral component acquires a non-vanishing vacuum-expectation value at a minimum of the potential, this therefore corresponds to the case of rank one of the matrix~$\twomat{K}^{(m)}$. 
 
It is convenient to separate these different cases corresponding to different rank conditions. We refer to the studies mentioned above for a detailed discussion of this subject in the 2HDM and the nHDM.  In analogy to the nHDM this can be done by imposing Lagrange multipliers ensuring the rank conditions. 
 %
\section{The Higgs multiplet potential}
 We now consider the general case of a multi-Higgs potential, where we have corresponding to each multiplicity~$m$ with $m=1, 2, 3, \ldots$ the number of $n_m$ copies of Higgs-boson multiplets.

In the potential we typically would like to consider only terms up to the fourth power of Higgs-boson fields, since otherwise, we would spoil renormalizability. However, with respect to the analysis here, we are not restricting the order of the potential.
Starting with $n_1$ copies of Higgs-boson singlets, there are, due to the singlet nature, no restrictions with respect to \sutwo{} symmetry for any term which may appear in the potential. 
As singlets have isospin zero, their hypercharge equals their electric charge. Therefore, gauge invariance imposes the restriction that each potential term must have a total electric charge of zero.
Concerning multiplets with multiplicity larger than one, we consider their irreducible totally symmetric tensor product representations. As we have seen, a multiplet of multiplicity~$m$ can be written in the symmetric representation with $m-1$ \sutwo ~indices, 
 $\Delta_{(i_1\ldots i_{m-1})}$. Writing all multiplets in this form
 allows us to build in a convenient way all gauge-invariant expressions which may appear in the potential. These multiplets transform in each index in the fundamental representation of~\sutwo. With the index notation borrowed from~\cite{Kannike:2023bfh} we write {\em conjugated} multiplets with upper indices,
 \begin{equation} \label{conjugated}
\Delta^{*i_1, \ldots, i_{m-1}} \equiv  (\Delta_{i_1, \ldots i_{m-1}})^*\qquad \text{with } i_k \in \{1,2\}\;.
\end{equation}
For instance, we write the anti quadruplet in the form
 $\Delta^{*(ijk)}$. The conjugated multiplets transform under \sutwo{} in every upper index in the anti-fundamental representation. Having the notation with upper and lower indices available, any product of multiplets with paired, that is, contracted indices, one above and one below, ensures invariance under \sutwo{} transformations. The hypercharge of the conjugated multiplets is also opposite to the original multiplets. An example, where we have one copy of a Higgs-boson doublet, is the potential of the Standard Model, where a quadratic term is conventionally written as $\varphi^\dagger \varphi$. In the symmetric representation we write this term $\Delta^{*i} \Delta_i$.
This term is manifestly gauge-invariant since the index~$i$ is contracted and any hypercharge assigned to the doublet appears once with positive and once with negative sign. Another example is the product $\Delta_i \Delta_j \Delta^{* ij}$ denoting an \sutwo-invariant product of two doublets with one triplet. 
It general we have to take care that the assignments of the hypercharges results in total hypercharge zero in every tensor product forming a term of the potential. 
For instance if we assign to the doublets 
$\Delta_i$ the hypercharge $1/2$ and to the triplet $\Delta_{(ij)}$ the hypercharge $+1$ we find the expression $\Delta_i \Delta_j \Delta^{* ij}$ to be gauge-invariant. 
We see at this point that we can form terms exclusively built from an odd number of copies of multiplets of the same multiplicity only if we assign to them a vanishing hypercharge. 

On the other hand, forming tensor products of different multiplicities, we can form odd  terms with non-vanishing hypercharges. An example is the term $\Delta_i \Delta_j \Delta^{* ij}$ with the appropriate assignments of hypercharge given in the last paragraph.
We can also form fundamental representations from anti-fundamental representations and vice versa, that is, we can raise and lower indices by applying the epsilon tensor (since for a  $\mathit{SU}(2)$ matrix we have $\epsilon U = U^* \epsilon$),
\begin{equation} \label{su2low}
\epsilon_{ij} \Delta^{* \ldots j \ldots }, \quad 
\epsilon^{ij} \Delta_{\ldots j \ldots }, \quad 
\text{with } \epsilon = i \sigma_2 = \begin{pmatrix} \phantom{+}0 & 1 \\ -1 & 0\end{pmatrix}\;.
\end{equation}
 The first expression of this form transforms with respect to the index~$i$ in the fundamental representation of~\sutwo{} and the second expression in the anti-fundamental representation.
An example is a term
$\Delta_j \Delta_k \Delta^{*m} \epsilon_{mi} \Delta^{* ijk}$ which is invariant under~\sutwo{}.

Eventually we arrive at the general recipe to construct a gauge invariant potential:

\begin{enumerate}
\item Write all multiplets beyond singlets as totally symmetric tensors, $\Delta_i$, $\Delta_{(ij)}$, $\Delta_{(ijk)}, \ldots$.  
\item Form conjugated multiplets with upper indices $\Delta^{*i}$, $\Delta^{* (ij)}$, 
$\Delta^{*(ijk)}, \ldots$.
\item Build products of the multiplet terms only with contracted indices. Here we consider that indices of the conjugated multiplets can be lowered and raised by multiplying with $\epsilon = i \sigma$. Singlets have no further restriction in this respect since they do not carry an~\sutwo{} index. 
\item We ensure that the assigned hypercharge in every tensor product adds to zero. 
\end{enumerate}

We now would like to illustrate this approach in some examples.
Let us again consider the potential of the Standard Model employing one copy of a Higgs-boson doublet 
$\varphi^{(2)}$. In the symmetric representation we write the doublet as $\Delta_i$. We form the conjugated doublet 
$\Delta^{* i}$ and find the well-known, most general potential up to order four,
\begin{equation}
V_{\text{SM}} = \mu^2  \Delta^{* i} \Delta_i + \lambda (\Delta^{* i} \Delta_i)^2\;.
\end{equation}
Next, let us consider the two-Higgs doublet model: In this case we have two copies of
Higgs-boson doublets which we write as $\Delta_{1\;i}$ and $\Delta_{2\;i}$ with the additional lower index labelling the number of the copy. The most general potential reads 
\begin{equation}
V_{\text{2HDM}} = 
\mu^2_{ab}  \Delta_{a}^{* i} \Delta_{b\; i} + 
\lambda_{abcd} \Delta_{a}^{* i} \Delta_{b\; i} \Delta_{c}^{* j} \Delta_{d\; j}\;.
\quad \text{with } a,b,c,d \in \{1, 2\}\;.
\end{equation}
Note that we have to sum over pairs of~\sutwo{} indices $i$, $j$, as well as over pairs of the copy labels~$a$, $b$, $c$, $d$.
Extending the range of the indices labelling the copies of the doublets, we can generalize the potential to cases with an arbitrary number of doublets. 

Eventually let us consider the case of one doublet corresponding to~$m=2$ with the assignment of hypercharge $+1/2$ and one triplet, that is $m=3$ with hypercharge $+1$. We find the most general gauge-invariant potential up to order four,
\begin{multline} \label{ex23}
V_{\text{23}} = 
\mu^2 \Delta^{* i} \Delta_{i} +  M^2 \Delta^{* ij} \Delta_{ij} + 
c_1 \Delta_{ij} \Delta^{* i} \Delta^{* j} +
c_1^*  \Delta^{* ij} \Delta_{i} \Delta_{j} + \\
\lambda_1 \Delta^{* i} \Delta_{i} \Delta^{* j} \Delta_{j} +
\lambda_2 \Delta^{* ij} \Delta_{ij} \Delta^{* kl} \Delta_{kl} +
\lambda_3 \Delta^{* ij} \Delta_{jk} \Delta^{* kl} \Delta_{li} +
\lambda_4 \Delta^{* ij} \Delta_{ij} \Delta^{* k} \Delta_k\;.
\end{multline}
It is instructive to compare a term $c_1 \Delta_{ij} \Delta^{* i} \Delta^{* j}$ mixing doublets with triplets to different notation. Often, the triplet is written as a Pauli vector in a basis $ \tilde{\tvec{\sigma}} = (\sigma_+, \sigma_3,  \sigma_-)^\trans$, with $\sigma_{\pm}=1/2 (\sigma_1\pm i \sigma_2)$. With the components of the triplet $\tilde{\phi} = (-\phi^1, \phi^2/\sqrt{2}, \phi^3)^\trans$ we have for the doublet and the triplet
\begin{equation} \label{term23conv}
\varphi_i = \Delta_i\;, \quad
\tilde{\Delta} =   \tilde{\phi}_a \tilde{\sigma}_a = \begin{pmatrix} \frac{\phi^2}{\sqrt{2}}  & -\phi^1 \\ \phi^3 & -  \frac{\phi^2}{\sqrt{2}} \end{pmatrix}.
\end{equation}
We see then that $(\Delta_{ij})= \tilde{\Delta} \epsilon$ indicating in particular that $(\Delta_{ij})$ and $\tilde{\Delta}$ transform differently under \sutwo. We eventually can write 
$\Delta^{* i}  \Delta_{ij} \Delta^{* j} = \varphi_i^*  \tilde{\phi}_a \tilde{\sigma}_a^{ij} \epsilon \varphi_j^*$. Often the field $\tilde{\varphi} = i \sigma_2 \varphi^* = \epsilon \varphi^*$ is defined emphasizing that $\tilde{\varphi}$ transforms in the fundamental representation. 

Let us note that we can also construct Higgs potential terms from explicit gauge-invariant bilinear terms~\eqref{bilinearsexp}. A general term formed from these bilinears of different multiplicities $m_1, \ldots, m_n$ can be written 
\begin{equation} \label{ximon0}
\xi^{\mu_1 \ldots \mu_n} K^{(m_1)}_{\mu_1} \dots K^{(m_n)}_{\mu_n}\;.
\end{equation}
Of course, any term formed from these bilinears in the potential has no restriction with respect to gauge invariance. Eventually we may also have mixed terms in the potential, out of Higgs multiplets in the symmetric representation accompanied by bilinears. Then the restrictions mentioned above for the symmetric doublet representations have to be taken into account. Let us also mention in this context that since the Higgs multiplets have always a symmetric representation, a description in terms of symmetric representations is exhaustive.  

\section{Change of basis}

Let us consider now a change of basis.  Basis changes are defined as a unitary mixing of the $n_m$ copies of Higgs multiplets in the fundamental representation
$\varphi^{(m)}_1,\ldots, \varphi^{(m)}_{n_m}$ with the same multiplicity $m$. 
We assume in this unitary mixing that the $n_m$ copies carry the same hypercharge. If this is not the case, the multiplets have to be separated with respect to sets carrying the same hypercharge (or any other quantum number) and therefore allow mixing. We have
\begin{equation} \label{basischange}
\psi'^{(m)} = \begin{pmatrix} {\varphi_1^{'(m)}}^\trans\\ \vdots\\ {\varphi_{n_m}^{'(m)}}^\trans \end{pmatrix}
=
U^{(m)} \begin{pmatrix} {\varphi_1^{(m)}}^\trans\\ \vdots\\ {\varphi_{n_m}^{(m)}}^\trans \end{pmatrix}
=
U^{(m)} \psi^{(m)} , \qquad \text{with } U^{(m)} \in (n_m \times n_m)\;.
\end{equation}
Under this unitary transformation the bilinear matrix transforms, as can be seen directly from its definition~\eqref{Kmatdef},
\begin{equation} \label{Kmatbasis}
\twomat{K}'^{(m)} = U^{(m)} \twomat{K}^{(m)} {U^{(m)}}^\dagger\;.
\end{equation}
This in turn means for the bilinears; see~\eqref{bilinearsexp},
\begin{multline} \label{Kbasis}
K'^{(m)}_\alpha = \tr ( \twomat{K}'^{(m)} \lambda_\alpha ) =
\tr (  \twomat{K}^{(m)} {U^{(m)}}^\dagger \lambda_\alpha U^{(m)})
=
R^{(m)}_{\alpha \beta} \tr ( \twomat{K}^{(m)} \lambda_\beta ) = R^{(m)}_{\alpha \beta} K^{(m)}_\beta\;,\\
\alpha, \beta \in \{0, \ldots, n_m^2-1\}\;,
\end{multline}
where we have written
\begin{equation} \label{contUR}
 {U^{(m)}}^\dagger \lambda_\alpha U^{(m)} = R^{(m)}_{\alpha \beta} \lambda_\beta\;.
 \end{equation}
Since $\lambda_0$ is the scaled identity matrix~\eqref{idscaled}, this reads,
 \begin{equation} \label{Kvbasis}
 K'^{(m)}_0 = K^{(m)}_0, \qquad 
 K'^{(m)}_a = R^{(m)}_{ab} K^{(m)}_b, 
 \qquad a, b \in \{1, \ldots, n_m^2-1\}. 
 \end{equation}
 The matrix $R^{(m)} = (R^{(m)}_{ab})$ has the properties
 \begin{equation}
 {R^{(m)}}^* = R^{(m)}, \qquad
 {R^{(m)}}^\trans R^{(m)} = \unitmatrix_{n_m^2-1},  \qquad
 \det (R^{(m)}) = 1\;.
 \end{equation}
 We see that $R^{(m)}(U^{(m)}) \subset SO(n_m^2-1)$ but only forms a 
 subset. Similar to the case of doublets we find for any type of multiplets, that a basis change corresponds in terms of bilinears to a proper rotation.
 
 Under basis changes, the symmetric representations of the multiplets $\Delta$, where we suppress the symmetric~\sutwo{} indices, transform in the same way as the multiplets in the (anti) fundamental representation, that is,
 \begin{equation} \label{basisDelta}
 \Delta^{(m)}_a \to 
 U^{(m)}_{ab}
 \Delta^{(m)}_b\;, \qquad
 \Delta^{(m)*}_c \to 
 \Delta^{(m)*}_d U^{(m)\dagger}_{dc}.
 \end{equation}
 The indices $a$, $b$, $c$, $d$ label the copies of the multiplets.
 Now suppose we deal with a model of different types of Higgs multiplets, for instance a model of two doublets and two triplets.
  We may consider the unitary mixing, i.~e., a change of basis, of the different multiplets separately. In the example the mixing of the two doublets could be considered independently from the mixing of the two triplets. 
  A Higgs potential is invariant under changes of bases if every gauge-invariant term is invariant. Let us consider a general term built from arbitrary multiplets, 
\begin{equation} \label{lambdamon}
\lambda^{a_1\ldots a_n c_1 \ldots c_m} 
\Delta^{(m_1)}_{a_1} \ldots \Delta^{(m_n)}_{a_n} 
\Delta^{*(m_{n+1})}_{c_1} \ldots \Delta^{*(m_{n+m})}_{c_m}.
\end{equation}
We emphasize that the multiplets can have different multiplicities indicated by the superscript index, whereas the lower indices $a_k$ and $c_k$ label the copy.   
This potential term is invariant under the changes of bases, \eqref{basisDelta}, if the corresponding parameter fulfills
\begin{equation} \label{lambdabasis}
\lambda^{a_1\ldots a_n c_1 \ldots c_m} 
=
U^{(m_{n+1})}_{c_1 d_1} \ldots U^{(m_{n+m})}_{c_m d_m}
\lambda^{b_1\ldots b_n d_1 \ldots d_m} 
U^{(m_1) \dagger}_{b_1 a_1} \ldots U^{(m_n) \dagger}_{b_n a_n}\;.
\end{equation}
 The unitary transformations appearing in this expression correspond to the copies of the different multiplets. 

Let us also consider terms of the potential written in terms of the adjoint representation, that is in terms of bilinears, which transform as shown in~\eqref{Kbasis}. Let us consider an arbitrary term written in terms of bilinears,
\begin{equation} \label{ximon}
\xi^{\mu_1 \ldots \mu_n} K^{(m_1)}_{\mu_1} \dots K^{(m_n)}_{\mu_n}\;.
\end{equation}
This term is invariant under changes of bases, if the corresponding parameter fulfills
\begin{equation} \label{xibasis}
 \xi^{\mu_1 \ldots \mu_n} =
  \xi^{\nu_1 \ldots \nu_n} 
  R^{(m_1) \trans}_{\nu_1 \mu_1} \ldots
  R^{(m_n) \trans}_{\nu_n \mu_n}\;.
  \end{equation} 
Also, we may have mixed terms in the potential written in terms of bilinears and symmetric representations. Then we find invariance if the corresponding parameters of the terms transform accordingly: Every copy index corresponding to a multiplet in the symmetric representation has to transform with a unitary transformation, and every index with respect to a bilinear with respect to an orthogonal rotation.

 \section{Symmetries of the Higgs-multiplet potential}
 
 Similar to the study of basis transformations we can find an appropriate treatment of symmetries of the Higgs-multiplet potential. Let us mention that this is a generalization of the study of symmetries in the 2HDM potential; see for instance~\cite{Ferreira:2010yh}. First of all we note that in a Higgs-multiplet model we have in particular to consider the kinetic terms of the Higgs multiplets. Therefore we have the condition that any symmetry transformation of the multiplets must be respected by the corresponding kinetic terms. As we have seen in the last section, basis transformations, that is, a unitary mixing among the multiplets of the same multiplicity keep the kinetic terms invariant. However, in addition to basis transformation we can simultaneously transform the space-time argument $x \to x'$ of the multiplets keeping the kinetic terms invariant. An example are CP transformations where in addition to a unitary transformation of the fields due to the charge transformation~(C), the parity transformation~(P) acts on the field arguments transforming $(\tvec{x},t) \to (-\tvec{x},t)$. 
 Therefore, for the $n_m$ copies of multiplets with multiplicity $m$ in the fundamental representation, the symmetry transformation extends the basis change transformations~\eqref{basischange} by a transformation of the argument, that is,
 \begin{equation} \label{symtrans}
\psi^{(m)}(x) \to \psi'^{(m)}(x) = 
U^{(m)} \psi^{(m)}(x') , \qquad \text{with } U^{(m)} \in (n_m \times n_m)\;.
\end{equation}
The most general transformation is therefore a unitary mixing given by the matrix $U^{(m)}$ accompanied by a transformation of the argument~$x \to x'$. 

Let us now consider the transformation of the Higgs multiplets in the symmetric representation.
Analogously to the transformation of multiplets in the fundamental representation we have to take the transformation of the argument into account; so we have in contrast to~\eqref{basisDelta},
\begin{equation} \label{symDelta}
 \Delta^{(m)}_a(x) \to 
 U^{(m)}_{ab}
 \Delta^{(m)}_b(x')\;, 
 \qquad
 \Delta^{(m)*}_c(x) \to 
 \Delta^{(m)*}_d(x') U^{(m)\dagger}_{dc}.
 \end{equation}
 Note that the indices $a$, $b$, $c$, $d$ label the copies of the multiplets.

The Higgs potential, formed from the Higgs multiplets in the symmetric representation is invariant under the transformations~\eqref{symDelta}, if every gauge-invariant term is invariant. Let us consider a general term of the potential build from arbitrary copies of multiplets in the symmetric representation, 
\begin{equation} \label{lambdamonsym}
\lambda^{a_1\ldots a_n c_1 \ldots c_m} 
\Delta^{(m_1)}_{a_1}(x) \ldots \Delta^{(m_n)}_{a_n}(x) 
\Delta^{*(m_{n+1})}_{c_1}(x) \ldots \Delta^{*(m_{n+m})}_{c_m}(x).
\end{equation}
This term in general will employ multiplets with different multiplicities indicated by the upper indices of the fields. We consider here the general case of~$n$ multiplets in the fundamental and $m$ multiplets in the anti-fundamental representation.
We see that a term~\eqref{lambdamonsym} is invariant under the symmetry transformation~\eqref{symDelta}, if the parameter fulfills the condition
\begin{equation} \label{lambdasym}
\lambda^{a_1\ldots a_n c_1 \ldots c_m} 
=
U^{(m_{n+1})}_{c_1 d_1} \ldots U^{(m_{n+m})}_{c_m d_m}
\lambda^{b_1\ldots b_n d_1 \ldots d_m} 
U^{(m_1) \dagger}_{b_1 a_1} \ldots U^{(m_n) \dagger}_{b_n a_n}\;.
\end{equation}

Eventually let us also consider bilinears in the potential with respect to possible symmetry transformations. We have seen in the last section that basis transformations of the bilinears are given by proper rotations of the bilinears~\eqref{Kbasis}. The rotation matrices were denoted by~$R^{(m)}$. However, considering general transformations keeping the kinetic terms of the bilinears invariant we are not restricted to proper rotations but may consider arbitrary orthogonal rotations. We shall denote these orthogonal rotations by~$X^{(m)}$ in order to distinguish them from the basis transformations $R^{(m)}$. Analogously to the previous discussion, we have to take also transformations of the argument $x \to x'$ of the bilinears into account. This is very similar to the case of symmetry transformations of $n_2$ Higgs-boson doublets~\cite{Maniatis:2015gma}.
This means we have for the bilinears to consider orthogonal transformations~$X^{(m)}$ and simultaneously transformations of the space-time argument $x \to x'$,
\begin{equation} \label{Xsymm}
K'^{(m)}_\alpha(x) = 
X^{(m)}_{\alpha \beta} K^{(m)}_\beta(x')
\qquad
\alpha, \beta \in \{0, \ldots, n_m^2-1\}\;.
\end{equation}
Generalizing~\eqref{Kvbasis}, any symmetry transformation should be of the form
 \begin{multline} \label{Xtrans}
  K^{(m)}_0(x)  \to K'^{(m)}_0(x) = K^{(m)}_0(x'), 
  \qquad 
    K^{(m)}_a(x) \to K^{'(m)}_a(x) = X^{(m)}_{ab} K^{(m)}_b(x'),\\
 X^{(m)} = (X^{(m)}_{ab}) \in O(n_m^2-1)\;, \qquad 
 a, b \in \{1, \ldots, n_m^2-1\}\;. 
 \end{multline}
 Note that for any symmetry transformation matrix $X^{(m)}$ of the bilinears, there is, except for a gauge transformation, a unique unitary transformation $U^{(m)}$; compare with~\eqref{contUR} and appendix~\ref{onetoone}, 
 \begin{equation} \label{contUX}
 {U^{(m)}}^\dagger \lambda_a U^{(m)} = X^{(m)}_{a b} \lambda_b\;.
 \end{equation}

Similar, we find the symmetry condition for a term of the potential written in terms of the adjoint representation, that is in terms of bilinears. Let us consider an arbitrary term written in terms of bilinears,
\begin{equation} \label{ximon2}
\xi^{\mu_1 \ldots \mu_n} K^{(m_1)}_{\mu_1}(x) \dots K^{(m_n)}_{\mu_n}(x)\;.
\end{equation}
This term is invariant under the transformation~\eqref{Xsymm}, if the 
bilinear parameter fulfills the condition
\begin{equation} \label{xibasisinv}
 \xi^{\mu_1 \ldots \mu_n} =
  \xi^{\nu_1 \ldots \nu_n} 
  X^{(m_1) \trans}_{\nu_1 \mu_1} \ldots
  X^{(m_n) \trans}_{\nu_n \mu_n}\;.
  \end{equation} 
Also, we may have mixed terms in the potential written in terms of bilinears and symmetric representations. Then the condition of invariance follow from the parameters: Under the transformation with respect to every index corresponding to a symmetric representation, and every index with respect to a bilinear the transformation of the parameter have to keep the term invariant. 

Let us turn to spontaneous symmetry breaking. This refers to the case of a symmetry of the potential, that is, we have a symmetry respected by the potential but not by the vacuum. In the case of a symmetric representation of the Higgs multiplets let us suppose we have a symmetry transformation~\eqref{symDelta} such that~\eqref{lambdasym} holds. Similar, for the case of potential terms written in bilinear form, we assume to have a transformation~\eqref{Xtrans} such that~\eqref{xibasisinv} holds (and similar for the mixed case). The symmetry is often called {\em explicit} in this context. For a vacuum of the potential we can then check whether this symmetry is respected. This requires for the multiplets in the symmetric representation,
\begin{equation} \label{sponDelta}
 \langle \Delta^{(m)}_a \rangle = 
 U^{(m)}_{ab}
  \langle \Delta^{(m)}_b \rangle
 ,
 \quad
\langle \Delta^{(m)*}_c \rangle =
 \langle \Delta^{(m)*}_d \rangle\; U^{(m)\dagger}_{dc},
 \qquad
 a,b,c,d \in \{1, \ldots, n_m\}\;.
 \end{equation}
For the case of bilinear representations we have to check at the vacuum,
 \begin{equation} \label{Xtransvac}
    \langle K^{(m)}_a \rangle = X^{(m)}_{ab} \langle K^{(m)}_b \rangle\;,
  \quad
 a,b \in \{1, \ldots, n_m^2-1\}\;.   
  \end{equation}
In the case that also~\eqref{sponDelta}, respectively~\eqref{Xtransvac} holds,
the vacuum respects the symmetry, that is, the symmetry is not spontaneously broken. 

\section{CP transformations in the Higgs multiplet model}
As an application of the symmetry study in the last section let us consider the standard CP transformation of the multiplets. The CP transformation is defined by charge conjugation of the multiplets simultaneously with a parity transformation of the space-time argument,
\begin{equation} \label{CPtrans}
\psi^{(m)}(x) \to \psi'^{(m)}(x) = {\psi^{(m)}}^*(x'), \qquad \text{with }
x=(t, \tvec{x}), \quad x'=(t, -\tvec{x})\;.
\end{equation}
Similar to the terminology in work~\cite{Maniatis:2007vn} we call these {\em standard} CP transformations in contrast to {\em generalized} CP transformations where simultaneously the multiplets are 
unitarily mixed\footnote{A generalized CP transformation is therefore given by, 
see~\eqref{symtrans}, $\psi^{(m)}(x) \to  U^{(m)} {\psi^{(m)}}^*(x')$}. 
Indeed it has been shown that the simultaneous mixing of the multiplets can give new types of transformations which do not correspond to standard CP transformations in a different basis.

The bilinear matrix~$\twomat{K}^{(m)}(x)$ transforms under the standard CP transformation as
\begin{equation} \label{KmatCP}
\twomat{K}^{(m)}(x) \to \twomat{K}^{'(m)}(x) = \psi'^{(m)}(x) {\psi'^{(m)}}^\dagger(x) =
(\psi^{(m)}(x') {\psi^{(m)}}^\dagger(x') )^\trans 
= {\twomat{K}^{(m)}}^\trans(x')\;.
\end{equation}
For the bilinears we get
\begin{multline} \label{KCP}
K^{(m)}_\alpha(x)  \to K^{'(m)}_\alpha(x) = \tr (\twomat{K}^{'(m)}(x) \lambda_\alpha)
= \tr (\twomat{K}^{(m)}(x') \lambda_\alpha^\trans) =
\tilde{C}^{(m)}_{\alpha \beta} K^{(m)}_\beta(x'), 
\\
\alpha, \beta \in \{0, \ldots, n_m^2-1\}\;,
\end{multline}
where we define the matrix~$\tilde{C}^{(m)}$ by
\begin{equation} \label{defC}
\lambda_\alpha^\trans = \tilde{C}^{(m)}_{\alpha \beta} \lambda_\beta\;.
\end{equation}
From the properties of the generalized Pauli matrices we find in particular,
\begin{equation}
\tilde{C}^{(m)}_{00} = 1\;,
\end{equation}
and all off-diagonal terms of $\tilde{C}^{(m)}$ vanishing. Therefore, we drop the trivial first row and the first column and define the matrix 
$C^{(m)}=(C^{(m)}_{a b})$ with $a, b \in \{1, \ldots, n_m^2-1\}$. 
Since the generalized Pauli matrices can be decomposed into symmetric and antisymmetric matrices, we get for the matrices $C^{(m)}$,
\begin{equation} \label{Cdiag}
C^{(m)} = \diag(\pm 1, \ldots, \pm 1)\;.
\end{equation}
The CP transformations of the bilinears read then,
\begin{multline} \label{CPKstand}
K^{(m)}_0(x) \to K^{(m)}_0(x’), \qquad
K^{(m)}_a(x) \to C^{(m)}_{ab} K^{(m)}_b(x’)\\
\text{with } 
\lambda_a^\trans = \tilde{C}^{(m)}_{ab} \lambda_b, 
\qquad
a, b \in \{1, \ldots, n_m^2-1\}\;.
\end{multline}
We see that, besides the space-time transformation of the argument of the multiplet fields, CP transformations give proper or improper rotations, depending on the number of symmetric and antisymmetric generalized Pauli matrices. For example, for the cases of two or three copies of multiplets of the same multiplicity we find reflection symmetry transformations, that is, matrices $C^{(m)}$ with determinant minus one. However, for instance in the case of four or five multiplets of the same multiplicity we can see that the standard CP transformations correspond to proper rotations in bilinear space.
 
 Eventually, the irreducible symmetric representations transform under standard CP transformations as
 \begin{equation} \label{CPDelta}
 \Delta^{(m)}_a(x) \to 
 \Delta^{*(m)}_a(x')\;, \qquad
 \Delta^{*(m)}_c(x) \to 
 \Delta^{(m)}_c(x').
 \end{equation}
 Note that we have suppressed again the \sutwo{} indices.
Let us mention that generalized CP transformations of the symmetric representations correspond to a unitary mixing - in addition to the complex conjugation and the parity transformation of the 
field argument\footnote{The generalized CP transformations are then
$ \Delta^{(m)}_a(x) \to  U^{(m)}_{ab}
 \Delta^{*(m)}_b(x')$, 
 $\Delta^{(m)*}_c(x) \to 
 \Delta^{(m)}_d(x') U^{(m)\dagger}_{dc}.$}.
Having formulated the standard CP transformation in terms of the bilinears, \eqref{CPKstand}, and in terms of the symmetric representations, \eqref{CPDelta}, we can verify the potential with respect to this symmetry. As discussed in the last section this requires to check the conditions~\eqref{lambdasym} in the case of the symmetric representation, 
respectively~\eqref{xibasisinv} in the case of bilinear representations. 

In the case we have a potential explicitly invariant under standard CP transformations, the conditions~\eqref{sponDelta}, respectively~\eqref{Xtransvac} determine whether this symmetry is respected by the vacuum.
Explicitly, these conditions for the standard CP symmetry considered here read for the symmetric representations
\begin{equation} \label{sponCPDelta}
 \langle \Delta^{(m)}_a \rangle = 
 \langle \Delta^{*(m)}_a \rangle\;.
 \end{equation}
In the case of bilinears we have the conditions
\begin{equation} \label{sponCPxi}
\langle K^{(m)}_a \rangle = C^{(m)}_{a b} \langle K^{(m)}_b \rangle\;.
\end{equation}
Of course, \eqref{sponCPDelta} and \eqref{sponCPxi} require to minimize the potential.
As an example let us consider the standard CP transformations in the case of $n_2$ copies of Higgs-boson doublets.
We write the potential in the symmetric representation,
\begin{equation}
V_{\text{NHDM}} = 
\mu^2_{ab}  \Delta_{a}^{* i} \Delta_{b\; i} + 
\lambda_{abcd} \Delta_{a}^{* i} \Delta_{b\; i} \Delta_{c}^{* j} \Delta_{d\; j}\;.
\quad \text{with } a,b,c,d \in \{1, \ldots n_2\}\;.
\end{equation}
The potential has to be real, requiring for the quadratic term that we must have ${\mu^2}^*_{ab} \Delta_{a\; i} \Delta_{b}^{* i} = 
\mu^2_{ab} \Delta_{a}^{* i} \Delta_{b\; i} = {\mu^2}^*_{ba} \Delta_{a}^{* i} \Delta_{b\; i}$, that is, 
$\mu^2_{ab} = {\mu^2}^*_{ba}$. Similar we have 
 $\lambda_{abcd} = \lambda_{badc}^*$. 
The CP transformation~\eqref{CPDelta} is given by $\Delta_{a\; i} \to \Delta_{a}^{* i}$ and
$\Delta_{a}^{* i} \to \Delta_{a\; i}$. Using the relation $\mu^2_{ab} = {\mu^2}^*_{ba}$ we find 
 $\mu^2_{ab} \Delta_{a\; i} \Delta_{b}^{* i} =
 {\mu^2}^*_{ab} \Delta_{a\; i} \Delta_{b}^{* i} =
 \mu^2_{ab} \Delta_{a\; i} \Delta_{b}^{* i}$, that is,
 the condition $\mu^2_{ab} = {\mu^2}^*_{ab}$ as expected. Similar we get the condition
$\lambda_{abcd} = \lambda_{abcd}^*$. 

Eventually, let us mention that for a transformation in form of a reflection it is often assumed that two transformations give a unity transformation. 
Interestingly, in~\cite{Ivanov:2015mwl} a three-Higgs-doublet model has been studied, where a generalized CP transformation gives unity only applied four times. In the corresponding potential, with the Higgs-boson doublets written in the fundamental representation, the parameters can not be all chosen to be real.

 \section{Example potential of two Higgs-boson doublets and three triplets}

Let us illustrate the formalism in an explicit example of a potential with two Higgs-boson doublets, 
\begin{equation}
\varphi_1^{(2)} = 
\begin{pmatrix} \phi_1^1 \\ \phi_1^2 \end{pmatrix}, \qquad
\varphi_2^{(2)} = 
\begin{pmatrix} \phi_2^1 \\ \phi_2^2 \end{pmatrix}\;.
\end{equation}
and three Higgs-boson triplets,
\begin{equation}
\varphi_i^{(3)} = 
\begin{pmatrix} \phi_i^{1} \\  \phi_i^2 \\ \phi_i^3 \end{pmatrix}, \qquad i \in \{1,2, 3\}\;.
\end{equation}
This corresponds to $n_2=2$ and $n_3=3$. We assume that the two Higgs-boson doublets carry the same hypercharge, as well as the three triplets carry the same hypercharge.
Usually we want the potential, as well as the complete Lagrangian of the model to be restricted by a symmetry. For a physically viable model this is typically required since, for instance, for two Higgs-boson doublets, the most general Yukawa couplings would lead to large flavor-changing neutral currents - which have never been observed. 
Here we will assume a specific potential of the form
\begin{multline} \label{Vex}
V_{\text{example}} = 
m_2^2 
\bigg({\varphi_1^{(2)}}^\dagger \varphi_1^{(2)}
+ {\varphi_2^{(2)}}^\dagger \varphi_2^{(2)} \bigg) 
+
m_3^2 
\bigg({\varphi_1^{(3)}}^\dagger \varphi_1^{(3)}
+ {\varphi_2^{(3)}}^\dagger \varphi_2^{(3)}  
+ {\varphi_3^{(3)}}^\dagger \varphi_3^{(3)} \bigg) 
\\
+ \lambda_2 
\bigg({\varphi_1^{(2)}}^\dagger \varphi_1^{(2)}
+ {\varphi_2^{(2)}}^\dagger \varphi_2^{(2)} 
\bigg)^2
+
\lambda_3
\bigg(
{\varphi_1^{(3)}}^\dagger \varphi_1^{(3)}
+ {\varphi_2^{(3)}}^\dagger \varphi_2^{(3)} 
+ {\varphi_3^{(3)}}^\dagger \varphi_3^{(3)} 
\bigg)^2
\\
+ \lambda_{23} 
({\varphi_1^{(2)}}^\dagger \varphi_2^{(2)} + {\varphi_2^{(2)}}^\dagger \varphi_1^{(2)})
\cdot 
({\varphi_1^{(3)}}^\dagger \varphi_2^{(3)} + {\varphi_2^{(3)}}^\dagger \varphi_1^{(3)})
\end{multline}
We translate the Higgs-doublet and triplet fields to the bilinear formalism. We form the $2 \times 2$ matrices $\psi^{(2)}$ and the $3 \times 3$ matrices $\psi^{(3)}$,
\begin{equation} \label{psiex}
\psi^{(2)} = \begin{pmatrix} {\varphi_1^{(2)}}^\trans\\ {\varphi_{2}^{(2)}}^\trans \end{pmatrix}, \qquad
\psi^{(3)} = \begin{pmatrix} 
{\varphi_1^{(3)}}^\trans\\ 
{\varphi_2^{(3)}}^\trans \\ 
{\varphi_3^{(3)}}^\trans 
\end{pmatrix}\;.
\end{equation}
Then we can built the bilinear $2 \times 2$, respectively, $3 \times 3$ matrices,
\begin{equation} \label{Kmatex}
\begin{split}
&\twomat{K}^{(2)} = \psi^{(2)} {\psi^{(2)}}^\dagger =
\begin{pmatrix} 
{\varphi_1^{(2)}}^\dagger \varphi_1^{(2)} &  {\varphi_2^{(2)}}^\dagger \varphi_1^{(2)} \\
{\varphi_1^{(2)}}^\dagger \varphi_2^{(2)} & {\varphi_{2}^{(2)}}^\dagger \varphi_{2}^{(2)}
\end{pmatrix},\\
&\twomat{K}^{(3)} = \psi^{(3)} {\psi^{(3)}}^\dagger =
\begin{pmatrix} 
{\varphi_1^{(3)}}^\dagger \varphi_1^{(3)} &  {\varphi_2^{(3)}}^\dagger \varphi_1^{(3)} &  {\varphi_3^{(3)}}^\dagger \varphi_1^{(3)}\\
{\varphi_1^{(3)}}^\dagger \varphi_2^{(3)} & {\varphi_2^{(3)}}^\dagger \varphi_{2}^{(3)} & {\varphi_3^{(3)}}^\dagger \varphi_2^{(3)} \\
{\varphi_1^{(3)}}^\dagger \varphi_3^{(3)} & {\varphi_2^{(3)}}^\dagger \varphi_3^{(3)} & {\varphi_3^{(3)}}^\dagger \varphi_3^{(3)} 
\end{pmatrix}\;.
\end{split}
\end{equation}
Now the bilinears follow in the basis of the Pauli and the Gell-Mann matrices together with the identity matrices $\sigma_0 = \unitmatrix_2$, and $\lambda_0 = \sqrt{2/3} \unitmatrix_3$,
\begin{equation} \label{biex}
 K^{(2)}_\alpha = \tr ( \twomat{K}^{(2)} \sigma_\alpha), 
 \quad \alpha \in \{0, \ldots, 3\},
 \quad
 K^{(3)}_\beta = \tr ( \twomat{K}^{(3)} \lambda_\beta), 
 \quad \beta \in \{0, \ldots, 8\}\;.
 \end{equation}
 Explicitly, we find 
\begin{alignat}{3} \label{Kexample2}
 &K^{(2)}_0 = {\varphi^{(2)}_1}^\dagger \varphi^{(2)}_1 + {\varphi^{(2)}_2}^\dagger \varphi^{(2)}_2, \qquad &&
 K^{(2)}_1 = {\varphi^{(2)}_1}^\dagger \varphi^{(2)}_2 + {\varphi^{(2)}_2}^\dagger \varphi^{(2)}_1, \nonumber\\
 &K^{(2)}_2 = i {\varphi_2^{(2)}}^\dagger \varphi^{(2)}_1 - i {\varphi^{(2)}_1}^\dagger \varphi^{(2)}_2,
 &&
 K^{(2)}_3 = {\varphi^{(2)}_1}^\dagger \varphi^{(2)}_1 - {\varphi^{(2)}_2}^\dagger \varphi^{(2)}_2\;,
 \\
\intertext{and}
 &K^{(3)}_0 = \sqrt{\frac{2}{3}} \left(
 {\varphi^{(3)}_1}^\dagger \varphi^{(3)}_1 + {\varphi^{(3)}_2}^\dagger \varphi^{(3)}_2 + {\varphi^{(3)}_3}^\dagger \varphi^{(3)}_3 \right),
  &&
 K^{(3)}_1 = {\varphi^{(3)}_1}^\dagger \varphi^{(3)}_2 + {\varphi^{(3)}_2}^\dagger \varphi^{(3)}_1, \nonumber\\
 &K^{(3)}_2 = {\varphi^{(3)}_1}^\dagger \varphi^{(3)}_3 + {\varphi^{(3)}_3}^\dagger \varphi^{(3)}_1, 
&&
 K^{(3)}_3 = {\varphi^{(3)}_2}^\dagger \varphi^{(3)}_3 + {\varphi^{(3)}_3}^\dagger \varphi^{(3)}_2, \nonumber\\
&K^{(3)}_4 = i {\varphi^{(3)}_2}^\dagger \varphi^{(3)}_1 - i {\varphi^{(3)}_1}^\dagger \varphi^{(3)}_2, 
&& K^{(3)}_5 = i {\varphi^{(3)}_3}^\dagger \varphi^{(3)}_1 - i {\varphi^{(3)}_1}^\dagger \varphi^{(3)}_3, \nonumber\\
 &K^{(3)}_6 = i {\varphi^{(3)}_3}^\dagger \varphi^{(3)}_2 - i {\varphi^{(3)}_2}^\dagger \varphi^{(3)}_3, 
&& K^{(3)}_7 = {\varphi^{(3)}_1}^\dagger \varphi^{(3)}_1 - {\varphi^{(3)}_2}^\dagger \varphi^{(3)}_2, \nonumber\\
&K^{(3)}_8 = \frac{1}{\sqrt{3}} \bigg(
 {\varphi^{(3)}_1}^\dagger \varphi^{(3)}_1 + 
 {\varphi^{(3)}_2}^\dagger \varphi^{(3)}_2 - 
 2 {\varphi^{(3)}_3}^\dagger \varphi^{(3)}_3 \bigg)\;.
  \end{alignat}
The inversion of these relations allow us to replace all the scalar products of the type ${\varphi^{(m)}_i}^\dagger \varphi^{(m)}_j$ by the bilinears,
\begin{align} \label{varphi2K2}
&{\varphi^{(2)}_1}^\dagger \varphi^{(2)}_1 = \frac{1}{2} \left(
K_0^{(2)} + K_3^{(2)} \right),\nonumber\\
&{\varphi^{(2)}_2}^\dagger \varphi^{(2)}_1 = \frac{1}{2} \left(
K_1^{(2)} - i K_2^{(2)} \right),
\qquad
&
{\varphi^{(2)}_2}^\dagger \varphi^{(2)}_2 = \frac{1}{2} \left(
K_0^{(2)} - K_3^{(2)} \right)\;,
\\
\intertext{and} \label{varphi2K3}
&{\varphi^{(3)}_1}^\dagger \varphi^{(3)}_1 = 
\frac{K_0^{(3)}}{\sqrt{6}} + \frac{K_7^{(3)}}{2} +\frac{K_8^{(3)}}{2\sqrt{3}}, \nonumber\\
&{\varphi^{(3)}_2}^\dagger \varphi^{(3)}_1 = \frac{1}{2} \left(
K_1^{(3)} - i K_4^{(3)} \right),
\qquad &
{\varphi^{(3)}_3}^\dagger \varphi^{(3)}_1 = \frac{1}{2} \left(
K_2^{(3)} - i K_5^{(3)} \right),\nonumber\\
&{\varphi^{(3)}_2}^\dagger \varphi^{(3)}_2 = 
\frac{K_0^{(3)}}{\sqrt{6}} - \frac{K_7^{(3)}}{2} +\frac{K_8^{(3)}}{2\sqrt{3}},
&
{\varphi^{(3)}_3}^\dagger \varphi^{(3)}_2 = \frac{1}{2} \left(
K_3^{(3)} - i K_5^{(3)} \right), \nonumber\\
&{\varphi^{(3)}_3}^\dagger \varphi^{(3)}_3 = 
\frac{K_0^{(3)}}{\sqrt{6}} - \frac{K_8^{(3)}}{\sqrt{3}}\;.
\end{align}
Note that we get the remaining expressions of the scalar products by recognizing that 
$({\varphi_i^{(m)}}^\dagger \varphi_j^{(m)})^* = {\varphi^{(m)}_j}^\dagger \varphi^{(m)}_i$ and recalling that the bilinears $K_\mu^{(m)}$ are real.

Now we express the potential~\eqref{Vex} in terms of bilinears by using~\eqref{varphi2K2}, 
\eqref{varphi2K3}, giving,
\begin{multline} \label{potbil}
V_{\text{example}} =
m_2^2 K^{(2)}_0 + \sqrt{\frac{3}{2}} m_3^2 K^{(3)}_0 
+ \lambda_2 (K^{(2)}_0)^2 
+ \frac{3}{2} \lambda_3 (K^{(3)}_0)^2 
+ \lambda_{23} K^{(2)}_1 K^{(3)}_1\;.
\end{multline}
Let us emphasize that this rather simple form of the potential arising in terms of bilinears comes from the fact that the potential is constructed solely from terms quadratic in doublets and triplets. In contrast, we could for instance consider a gauge-invariant term in the potential which couples one triplet to a pair of doublets, similar to the example shown in~\eqref{ex23}. 

We study the potential with respect to standard CP symmetries. From~\eqref{CPKstand} 
we find the explicit transformation matrices for the two doublets and the three triplets,
\begin{equation} \label{CPexp}
C^{(2)} = \diag(1, -1, 1), \qquad
C^{(3)} = \diag(1,1, -1,-1,-1,1, 1)\;.
\end{equation}
We can now check the conditions of the parameters, \eqref{xibasisinv}, to see whether or not the potential is explicitly CP conserving. In particular we find the only non-trivial condition
$\lambda_{23} C^{(2)}_{11} C^{(3)}_{11} = \lambda_{23}$. With view at~\eqref{CPexp} this condition is fulfilled, therefore, the potential is explicitly symmetric under CP transformations.

 \section{Conclusions}
 \label{conclusions}
 
There is substantial motivation to explore models with Higgs multiplets beyond singlets and doublets. 
Despite the stringent constraints imposed by the electroweak $\rho$ parameter, viable scenarios still exist where Higgs bosons with higher multiplicities can play a role. 

Building on our previous work, we have demonstrated that gauge redundancies in the description of Higgs multiplets generally obscure the physical interpretation of these models. For an arbitrary number of Higgs-boson multiplets, we have employed gauge-invariant methods to determine the domain, utilizing bilinear representations - similar to the approach developed for multiple Higgs-boson doublets.

However, the potential may include terms that, while gauge-invariant, are not bilinear in all the present multiplets. Using methods akin to those discussed in~\cite{Kannike:2023bfh}, we treated the multiplets within the irreducible symmetric representation, with \sutwo{} indices. Just as Lorentz invariance is reflected by contracted indices, contracted \sutwo{} indices signify \sutwo{} invariance. This approach allows for the construction of all gauge-invariant potential terms for an arbitrary assignment of hypercharge to the multiplets.

Finally, we have investigated general symmetries in terms of bilinears and symmetric representations. Specifically, we illustrated how standard CP transformations manifest in multi-Higgs potentials and provided an explicit example to demonstrate this.

%
 \acknowledgments
 We are very thankful to Otto Nachtmann, Lohan Sartore, and Kristjan Kannike for valuable comments and suggestions.
 This work is supported by Chile ANID FONDECYT project 1200641. 
 
 \appendix
 
 \section{One-to-one map of bilinears to Higgs multiplets}
 \label{onetoone}
Here we want to show the one-to-one correspondence of the Higgs bilinears to the fundamental Higgs multiplets - except for unphysical gauge degrees of freedom. The proof of this is a generalization of the proof given in theorem~5 of~\cite{Maniatis:2006fs}.
We consider the general case of $n_m$ copies of Higgs multiplets $\varphi^{(m)}_1(x), \ldots,  \varphi^{(m)}_{n_m}(x)$. As outlined above these $n_m$ multiplets of the same type are written together in one matrix,
\begin{equation} \label{psigen}
\psi^{(m)} = 
\begin{pmatrix}
{\varphi_1^{(m)}}^\trans\\
\vdots\\
{\varphi_{n_m}^{(m)}}^\trans
\end{pmatrix} 
=
\begin{pmatrix} 
\phi^{1}_1& \cdots & \phi^m_1\\
\vdots & \vdots\\
\phi^{1}_{n_m}& \cdots & \phi^m_{n_m}
\end{pmatrix}
\qquad \in (n_m \times m)\;.
\end{equation}

We shall proof now that for any hermitean, positive semidefinite matrix $\twomat{K}^{(m)}$ of dimension $n_m \times n_m$ of rank less than or equal to $m$, there are Higgs multiplets $\varphi^{(m)}_1(x), \ldots,  \varphi^{(m)}_{n_m}(x)$ such that 
$\twomat{K}^{(m)} = \psi^{(m)} {\psi^{(m)}}^\dagger$. For a given matrix $\twomat{K}^{(m)}$ of dimension $n_m \times n_m$ the Higgs multiplets are determined uniquely - except for a electroweak gauge transformation~\eweakgroup.\\

Suppose we have a hermitean, positive~semidefinite matrix $\twomat{K}^{(m)}$ of dimension $n_m \times n_m$ of rank less than or equal to $m$. We can diagonalize this matrix such that
\begin{equation} \label{twomatdiag}
\twomat{K}^{(m)} = W(x)
\diag( \kappa_1(x), \ldots, \kappa_m(x), 0, \ldots, 0) W^\dagger(x)\;
\end{equation}
The matrices $W(x)$ are unitary matrices of dimension $(n_m \times n_m)$ and the eigenvalues fulfill
\begin{equation} \label{kappaeigen}
\kappa_i(x) \ge 0, \qquad i \in \{1, \ldots, m\}\;.
\end{equation}
Having diagonalized $\twomat{K}^{(m)}$ we set 
\begin{equation}
\psi^{(m)}(x) = W(x) 
\left(
\begin{array}{ccc} 
\sqrt{\kappa_1(x)} & & 0\\
0 & \ddots &  0\\
0 & & \sqrt{\kappa_m(x)}\\
\hline
 & 0 &\\
\end{array}
\right).
\end{equation}
We see that $\psi^{(m)}(x)$ has the form~\eqref{psigen} and that $\twomat{K}^{(m)}(x) = \psi^{(m)} (x) {\psi^{(m)}}^\dagger(x)$.

It remains to show that the $\psi^{(m)}(x)$ fulfilling $\twomat{K}^{(m)}(x) = \psi^{(m)}(x) {\psi^{(m)}}^\dagger(x)$ are unique up to gauge transformations. Suppose, we have two field configurations $\psi^{(m)}(x)$ and ${\psi^{(m)}}'(x)$ of the multiplets giving the same matrix
\begin{equation} \label{twomatphiphip}
\twomat{K}^{(m)}(x) = \psi^{(m)}(x) {\psi^{(m)}}^\dagger(x) = {\psi^{(m)}}'(x) {{\psi^{(m)}}'}^\dagger(x)
\end{equation}
Diagonalization of $\twomat{K}^{(m)}$ as in~\eqref{twomatdiag} then reads
\begin{equation}
\left( \begin{array}{c|c}
\begin{array}{ccc}
 \kappa_1(x) &  & 0\\
 &\ddots &\\
 0 & &  \kappa_m(x)
 \end{array}
 & 0 \\
 \hline
 0 & 0 
 \end{array} \right)
  =
\left( W^\dagger \psi^{(m)} \right) \left( W^\dagger \psi^{(m)} \right)^\dagger =
\left( W^\dagger {\psi^{(m)}}' \right) \left( W^\dagger {\psi^{(m)}}' \right)^\dagger\;.
\end{equation}
From this follows that we can write 
\begin{equation}
\left( W^\dagger \psi^{(m)} \right) = 
\left(
\begin{array}{c}
\chi^{(m)}_1(x)^\trans\\
\vdots\\
\chi^{(m)}_{n_m}(x)^\trans\\
\hline
0
\end{array}
\right),
\qquad
\left( W^\dagger {\psi^{(m)}}' \right) = 
\left(\begin{array}{c}
\chi'^{(m)}_1(x)^\trans\\
\vdots\\
\chi'^{(m)}_{n_m}(x)^\trans\\
\hline
0
\end{array} \right),
\end{equation}
where
\begin{equation}
{\chi^{(m)}_i}^\dagger(x) \chi^{(m)}_j(x) =
{\chi'^{(m)}_i}^\dagger(x) \chi'^{(m)}_j(x) = \kappa_i(x) \delta_{ij}\;.
\end{equation}
From this we conclude that there is a unitary matrix $U_G \in U(m \times m)$ such that
 \begin{equation}
 \chi'^{(m)}_i = U_G\; \chi^{(m)}_i, \qquad i \in \{1, \ldots m\}\;.
 \end{equation}
 With this we find for the multiplets ${\psi^{(m)}}'$ and $\psi^{(m)}$
 \begin{equation}
 W^\dagger(x) {\psi^{(m)}}'(x) = W^\dagger(x) \psi^{(m)}(x) U_G^\trans (x),
 \end{equation}
 and noting that $W(x)$ is unitary we find eventually
 \begin{equation}
 {\psi^{(m)}}'(x) = \psi^{(m)}(x) U_G^\trans (x),
 \end{equation}
 that is, as stated, $ {\psi^{(m)}}'(x)$ and $\psi^{(m)}(x)$ are related by a gauge transformation.
 
\bibliographystyle{JHEP}
\bibliography{references}
\end{document}